\RequirePackage{fixltx2e}
\documentclass[a4paper]{isp}
\usepackage[american]{babel}
\usepackage{graphicx}
\usepackage{amsmath,amsthm,mathtools}
\usepackage{url}
\usepackage[capitalise,nameinlink]{cleveref}
\usepackage{xcolor}
\def\aap{A\&A\,  }%% Astronomy and Astrophysics

\def\apj{ApJ\,  }%% Astrophysical Journal
\def\apjs{ApJS  }%% Astrophysical Journal, Supplement
\def\mnras{MNRAS\,  }%% Monthly Notices of the RAS

\def\prl{Phys. Rev. Lett.    }% % Physical Review Letters 
\begin{document}
\pdfgentounicode=1
\title
{
On the shape of the local bubble
}
\author{Lorenzo Zaninetti}
\institute{
Physics Department,
 via P.Giuria 1, I-10125 Turin,Italy \\
 \email{zaninetti@ph.unito.it}
}

\maketitle

\begin {abstract}
The shape of the  local bubble  is  modeled in the framework
of the thin layer approximation.
The asymmetric shape of the  local bubble  is  simulated 
by introducing axial profiles for the density of the interstellar medium,
such as 
exponential,
Gaussian, 
inverse square dependence 
and
Navarro--Frenk--White.
The availability  of some observed asymmetric profiles for 
the local bubble 
allows us to match theory and observations via 
the observational
percentage of reliability.
The model is compatible with the presence of radioisotopes on Earth.
\end{abstract}

\section{Introduction}

The local bubble (LB) is a   low-density region  that surrounds the Sun.
Because it is emitting in the X-rays, it
is also called Local Hot Bubble (LHB), see \cite{Arnaud1984,Slavin2016}.
In the framework of  thermal
ionization equilibrium,
the temperature is
$kT = (0.097 \pm 0.013) keV$  or  $ T= (1.1252 \pm 0.15)\,10^6 K$, 
see \cite{Liu2017}.
Recently, the following features of the LB have been discussed:
the variations of the polarization degree P, see \cite{Gontcharov2019}; 
and
the  polarization from the interstellar medium,
due to irregular dust grains aligned with the magnetic
field, see \cite{Medan2019}.
The presence of $^{60}$Fe  in deep-sea measurements on Earth 
has triggered the study of the LB-sun interaction, see \cite{Knie2004}. 
We now select some theoretical efforts that
model the LB, as follows: the 
one-dimensional hydrocode  with non-equilibrium ion evolution and dust,
see \cite{Smith1998,Smith2001};
different tests to explain the FUSE data,
see \cite{Welsh2002,Breitschwerdt2009}; 
the use of the
parallel adaptive mesh refinement code EAF-PAMR,
see \cite{Deavillez2012};
hydrodynamical simulations of the LB, see \cite{Feige2018};
and
the study of the 3D structure of the  magnetic field, see \cite{Alves2018}.

These models leave some questions
unanswered, or only partially answered, as follows:
\begin{itemize}
\item  Can we model  the LB in the framework of the thin 
       layer expansion of a shell in 
       an   interstellar medium (ISM)  with symmetry 
       in respect to the equatorial plane of the explosion? 
\item  Can we compare the data of the  theoretical expansion, which 
       is a function of the latitude, with the observed 
       profiles of expansion of the LB?
\item  What  is the range of reliability of the 
       Taylor expansion and Pad\'e approximation of the 
       theoretical expansion in the framework of the thin layer?
\item  Can we model the LB-Sun interaction?
\end{itemize}
To answer these questions:
Section \ref{section_density} analyzes
four   profiles of density for the interstellar medium (ISM);
Section \ref{section_motion} derives four equations 
of motion for the LB;
and  
Section \ref{section_astro}  discusses 
the results for the four   equations of motion
in terms of reliability of the model,
it also introduces the interaction of many bubbles, 
discusses the $^{60}$Fe-signature    
and explores the interaction of many bubbles.

\section{The density profiles}

\label{section_density}
A point in  Cartesian coordinates is characterized by
$x,y$ and $z$,  
and  the position of the origin
is the center of the LB.
The same point in spherical coordinates is characterized by
the radial distance $r \in[0,\infty]$,
the polar angle     $\theta \in [0,\pi]$,
and the azimuthal angle $\varphi \in [0,2\pi]$.

The following profiles are considered:
exponential,
Gaussian, 
inverse square dependence 
and
Navarro--Frenk--White.  

\subsection{An exponential profile}

The density  is  assumed to have the following
exponential dependence on $z$
in Cartesian coordinates:
\begin{equation}
 \rho(z;b,\rho_0) =
\rho_0  \exp{(-z/b)}
\quad ,
\label{profexponential}
\end{equation}
where $b$ represents the scale.
In spherical coordinates,
the density has the following piecewise dependence
\begin{equation}
 \rho(r;r_0,b,\rho_0) =
  \begin{cases}
    \rho_0                                 & \quad \text{if } r    \leq r_0\\
    \rho_0 \exp{(-\frac{r\cos(\theta)}{b})}  & \quad \text{if } r    >   r_0\\
  \end{cases}
  \label{profexponentialr}
\quad ,
\end{equation}
which has a jump discontinuity at $r=r_0$ when $\theta>0$.
Given a solid angle  $\Delta \Omega$,
the total mass swept,   $M(r;r_0,b,\theta,\rho_0,\Delta \Omega) $,
in the interval $[0,r]$ is
\begin{multline}
M (r;r_0,b,\theta,\rho_0,\Delta \Omega)= \\
\Bigg (
\frac{1}{3}\,\rho_{{0}}{r_{{0}}}^{3}-{\frac {b \left( {r}^{2} \left( \cos
 \left( \theta \right)  \right) ^{2}+2\,rb\cos \left( \theta \right) +
2\,{b}^{2} \right) \rho_{{0}}}{ \left( \cos \left( \theta \right) 
 \right) ^{3}}{{\rm e}^{-{\frac {r\cos \left( \theta \right) }{b}}}}}
\\
+
{\frac {b \left( {r_{{0}}}^{2} \left( \cos \left( \theta \right) 
 \right) ^{2}
+2\,r_{{0}}b\cos \left( \theta \right) +2\,{b}^{2}
 \right) \rho_{{0}}}{ \left( \cos \left( \theta \right)  \right) ^{3}}
{{\rm e}^{-{\frac {r_{{0}}\cos \left( \theta \right) }{b}}}}}
\Bigg )
\Delta \Omega
\quad .
\label{massexponential}
\end{multline}

\subsection{A  Gaussian profile}

The density  is  assumed to have the following Gaussian
dependence on $z$
in Cartesian coordinates:
\begin{equation}
 \rho(z;b,\rho_0) =
\rho_0  {{\rm e}^{-\frac{1}{2}\,{\frac {{z}^{2}}{{b}^{2}}}}}
\quad ,
\label{profgaussianzero}
\end{equation}
where $b$ represents a parameter.
In spherical coordinates,
the density is
\begin{equation}
 \rho(r;r_0,b,\rho_0) =
  \begin{cases}
    \rho_0                                               & \quad \text{if } r    \leq r_0\\
    \rho_0 {{\rm e}^{-\frac{1}{2}\,{\frac {{z}^{2}}{{b}^{2}}}}}  & \quad \text{if } r    >   r_0\\
  \end{cases}
  \label{profgaussian}
\quad ,
\end{equation}
and presents   a jump discontinuity at $r=r_0$
when $\theta>0$.
The total mass swept,   $M(r;r_0,b,\theta,\rho_0) $,
in the interval $[0,r]$
is
\begin{multline}
M (r;r_0,b,\theta,\rho_0)=
\Bigg (
\frac{1}{3}\,\rho_{{0}}{r_{{0}}}^{3}
+\rho_{{0}} \bigg( -{\frac {r{b}^{2}}{
 \left( \cos \left( \theta \right)  \right) ^{2}}{{\rm e}^{-\frac{1}{2}\,{
\frac {{r}^{2} \left( \cos \left( \theta \right)  \right) ^{2}}{{b}^{2
}}}}}}
\\
+\frac{1}{2}\,{\frac {{b}^{3}\sqrt {\pi}\sqrt {2}}{ \left( \cos \left( 
\theta \right)  \right) ^{3}}{\rm erf} \left(\frac{1}{2}\,{\frac {\sqrt {2}
\cos \left( \theta \right) r}{b}}\right)} \bigg ) 
-\rho_{{0}} \bigg( -
{\frac {r_{{0}}{b}^{2}}{ \left( \cos \left( \theta \right)  \right) ^{
2}}{{\rm e}^{-\frac{1}{2}\,{\frac {{r_{{0}}}^{2} \left( \cos \left( \theta
 \right)  \right) ^{2}}{{b}^{2}}}}}}
\\
+\frac{1}{2}\,{\frac {{b}^{3}\sqrt {\pi}
\sqrt {2}}{ \left( \cos \left( \theta \right)  \right) ^{3}}{\rm erf} 
\left(\frac{1}{2}\,{\frac {\sqrt {2}\cos \left( \theta \right) r_{{0}}}{b}}
\right)} \bigg ) 
\Bigg ) 
 \Delta \Omega
\quad ,
\label{massgaussian}
\end{multline}

\noindent 
where $\mathop{\mathrm{erf}}(x)$
is the error function, defined by
\begin{equation}
\mathop{\mathrm{erf}\/}\nolimits
(x)=\frac{2}{\sqrt{\pi}}\int_{0}^{x}e^{-t^{2}}dt
\quad .
\end{equation}

\subsection{The inverse square dependence}

The density  is  assumed to have the following
dependence on $z$
in Cartesian coordinates,
\begin{equation}
 \rho(z;z_0,\rho_0) =\rho_0\left( 1+{\frac {z}{{\it z_0}}} \right) ^{-2}
\quad .
\label{squareprofile}
\end{equation}
In this paper, we will adopt the following 
density profile  in spherical coordinates
\begin{equation}
 \rho(r;r_0,b,\rho_0) =
  \begin{cases}
    \rho_0                                               & \quad \text{if } r    \leq r_0\\
    \rho_0\left( 1+{\frac {r\,\cos(\theta)}{{\it z_0}}} \right) ^{-2}  & \quad \text{if } r    >   r_0\\
  \end{cases}
  \label{profsquare}
\quad 
\end{equation}
where the parameter $z_0$ fixes the scale and  $\rho_0$ is the
density at $z=z_0$.
The above density presents    a jump discontinuity at $r=r_0$
when $\theta>0$.
The mass $M_0$ swept
in the interval $[0,r_0]$
is
\begin{equation}
M_0 =
\frac{1}{3}\,\rho_{{0}}\,{r_{{0}}}^{3} \,\Delta \Omega
\quad .
\end{equation}
The total mass swept, $M(r;r_0,z_0,\theta,\rho_0,\Delta \Omega) $,
in the interval $[0,r]$
is
\begin{eqnarray}
M(r;r_0,z_0,\theta,\rho_0,\Delta \Omega)=
\nonumber \\ 
\Biggr (
\frac{1}{3}\,\rho_{{0}}{r_{{0}}}^{3}+{\frac {\rho_{{0}}{b}^{2}r}{
 \left( \cos \left( \theta \right)  \right) ^{2}}}-2\,{\frac {\rho_{{0
}}{b}^{3}\ln  \left( r\cos \left( \theta \right) +b
 \right) }{ \left( \cos \left( \theta \right)  \right) ^{3}}}
\nonumber \\
-{\frac {
\rho_{{0}}{b}^{4}}{ \left( \cos \left( \theta \right)  \right) ^
{3} \left( r\cos \left( \theta \right) +b \right) }}-{\frac {
\rho_{{0}}{b}^{2}r_{{0}}}{ \left( \cos \left( \theta \right) 
 \right) ^{2}}}+2\,{\frac {\rho_{{0}}{b}^{3}\ln  \left( r_{{0}}
\cos \left( \theta \right) +b \right) }{ \left( \cos \left( 
\theta \right)  \right) ^{3}}}
\nonumber  \\
+{\frac {\rho_{{0}}{b}^{4}}{
 \left( \cos \left( \theta \right)  \right) ^{3} \left( r_{{0}}\cos
 \left( \theta \right) +b \right) }}
\Biggl )   
\Delta \Omega
\quad .
\label{mass_square}
\end{eqnarray}

\subsection{Navarro--Frenk--White profile}

The usual Navarro--Frenk--White (NFW) distribution  has a 
dependence on $r$
in spherical  coordinates  of the type
\begin{equation}
 \rho(r;r_0,b,\rho_0) =
\frac
             {
              \rho_{{0}}r_{{0}} \left( b+r_{{0}} \right) ^{2}
             }
             {
              r \left( b+r \right) ^{2}
             }
\quad ,
\label{profilenfwr}
\end{equation}
where $b$ represents the scale,
see \cite{Navarro1996} 
for more details.
The NFW profile along the axis $z$ can be obtained by
substituting $r$ with $r \,\cos(\theta)=z$ 
\begin{equation}
 \rho(r;r_0,b,\rho_0,\theta) =
\frac
             {
              \rho_{{0}}r_{{0}} \left( b+r_{{0}} \right) ^{2}
             }
             {
              r\,\cos(\theta) \left( b+r\,\cos(\theta) \right) ^{2}
             }
\quad ,
\label{profilenfwz}
\end{equation}

The piece-wise  density is
\begin{equation}
 \rho (r;r_0,b,\rho_0\theta)  = \left\{ \begin{array}{ll} 
            \rho_0                      & \mbox         {if $r \leq r_0 $ } \\
            \frac
             {
              \rho_{{0}}r_{{0}} \left( b+r_{{0}} \right) ^{2}
             }
             {
              r\,\cos(\theta) \left( b+r\,\cos(\theta) \right) ^{2}
             }
   & \mbox {if $r >    r_0 $ } 
            \end{array}
            \right.
\label{piecewisenfw}
\end{equation}
and  has  a jump discontinuity at $r=r_0$ when $\theta>0$.
The total mass swept,   $M(r;r_0,b,\rho_0\theta) $,
in the interval $[0,r]$ is
\begin{eqnarray}
M (r;r_0,b,\theta,\rho_0,\Delta \Omega)= \nonumber \\
=
\Big (
\frac{1}{3}\,\rho_{{0}}{r_{{0}}}^{3}+{\frac {\rho_{{0}} \left(  \left( b+r\cos
 \left( \theta \right)  \right) \ln  \left( b+r\cos \left( \theta
 \right)  \right) +b \right)  \left( b+r_{{0}} \right) ^{2}r_{{0}}}{
 \left( \cos \left( \theta \right)  \right) ^{3} \left( b+r\cos
 \left( \theta \right)  \right) }}
\nonumber  \\
-{\frac {\rho_{{0}} \left(  \left( b
+r_{{0}}\cos \left( \theta \right)  \right) \ln  \left( b+r_{{0}}\cos
 \left( \theta \right)  \right) +b \right)  \left( b+r_{{0}} \right) ^
{2}r_{{0}}}{ \left( \cos \left( \theta \right)  \right) ^{3} \left( b+
r_{{0}}\cos \left( \theta \right)  \right) }}
\Bigg )
\Delta \Omega
\end{eqnarray}

\section{The  thin layer approximation}
\label{section_motion}

The conservation of the momentum in
spherical coordinates
along  the  solid angle  $\Delta \Omega$
in the framework of the thin
layer approximation  states that
\begin{equation}
M_0(r_0) \,v_0 = M(r)\,v
\quad ,
\end{equation}
where $M_0(r_0)$ and $M(r)$ are the swept masses at $r_0$ and $r$,
and $v_0$ and $v$ are the velocities of the thin layer at $r_0$ and $r$.
This conservation law can be expressed as a differential equation
of the first order by inserting $v=\frac{dr}{dt}$:
\begin{equation}
M(r)\, \frac{dr}{dt} = M_0(r_0)\, v_0
\quad .
\end{equation}
In the first phase from $r=0$ to $r=r_0$  the density is 
constant and the explosion is symmetrical.
In the second phase the density is function of the polar angle
$\theta$ and therefore the shape of the advancing expansion
is asymmetrical.
The equation of motion for the four profiles is now derived.

\subsection{Motion with a constant density}

In the case of constant density of the ISM,  
the analytical solution 
for the trajectory is 
\begin{equation} 
r(t;t_0,r_0,v_0) =
\sqrt [4]{4\,{r_{{0}}}^{3}v_{{0}} \left( t-t_{{0}} \right) +{r_{{0}}}^
{4}}
\label{rtconstant}
\quad , 
\end {equation}
and the velocity is 
\begin{equation} 
v(t;t_0,r_0,v_0) =
\frac
{
{r_{{0}}}^{3}v_{{0}}
}
{
 \left( 4\,{r_{{0}}}^{3}v_{{0}} \left( t-t_{{0}} \right) +{r_{{0}}}^{4
} \right) ^{3/4}
}
\label{vtconstant}
\quad , 
\end {equation}
where   $r_0$ and $v_0$ are the position and the velocity
when    $t=t_0$,
see \cite{Dyson1997,Padmanabhan_II_2001}.

\subsection{Motion with an exponential profile}

In the case of  an exponential density profile
for the ISM,
as given by equation (\ref{profexponentialr}),
the differential equation
that models momentum conservation
is
\begin{multline}
 \Bigg ( \frac{1}{3}\, {r_{{0}}}^{3}-{\frac {b \left(  \left( r
 \left( t \right)  \right) ^{2} \left( \cos \left( \theta \right) 
 \right) ^{2}+2\,r \left( t \right) b\cos \left( \theta \right) 
+2\,{b
}^{2} \right)  }{ \left( \cos \left( \theta \right)  \right) 
^{3}}{{\rm e}^{-{\frac {\cos \left( \theta \right) r \left( t \right) 
}{b}}}}}
\\
+{\frac {b \left( {r_{{0}}}^{2} \left( \cos \left( \theta
 \right)  \right) ^{2}+2\,r_{{0}}b\cos \left( \theta \right) 
+2\,{b}^{
2} \right)  }{ \left( \cos \left( \theta \right)  \right) ^{3
}}{{\rm e}^{-{\frac {r_{{0}}\cos \left( \theta \right) }{b}}}}}
 \Bigg ) {\frac {\rm d}{{\rm d}t}}r \left( t \right) =\frac{1}{3}\, {
r_{{0}}}^{3}v_{{0}}
\quad .
\label{eqndiffexp}
\end{multline}
There is no analytical solution.

\subsection{Motion with a Gaussian profile}

In the case of  a Gaussian  density profile
for the ISM,
as given by equation (\ref{profgaussian}),
the differential equation
that models momentum conservation
is
\begin{eqnarray}
\Bigg  ( \frac{1}{3}\,\rho_{{0}}{r_{{0}}}^{3}+\rho_{{0}}  \Big ( -{\frac {r
   ( t   ) {b}^{2}}{   ( \cos   ( \theta   ) 
   ) ^{2}}{{\rm e}^{-\frac{1}{2}\,{\frac {   ( r   ( t   ) 
   ) ^{2}   ( \cos   ( \theta   )    ) ^{2}}{{b}^{2}
}}}}}+\frac{1}{2}\,{\frac {{b}^{3}\sqrt {\pi}\sqrt {2}}{   ( \cos   ( 
\theta   )    ) ^{3}}{\rm erf} \big  (\frac{1}{2}\,{\frac {\sqrt {2}
\cos   ( \theta   ) r   ( t   ) }{b}} \big )}  \Big ) 
\nonumber \\
-
\rho_{{0}} \Big  ( -{\frac {r_{{0}}{b}^{2}}{   ( \cos   ( \theta
   )    ) ^{2}}{{\rm e}^{-\frac{1}{2}\,{\frac {{r_{{0}}}^{2}   ( 
\cos   ( \theta   )    ) ^{2}}{{b}^{2}}}}}}+\frac{1}{2}\,{\frac {{b
}^{3}\sqrt {\pi}\sqrt {2}}{   ( \cos   ( \theta   ) 
   ) ^{3}}{\rm erf} \big  (\frac{1}{2}\,{\frac {\sqrt {2}\cos   ( \theta
   ) r_{{0}}}{b}} \big )} \Big  ) \Bigg   ) {\frac {\rm d}{{\rm d}t
}}r   ( t   ) =\frac{1}{3}\,\rho_{{0}}{r_{{0}}}^{3}v_{{0}}
\quad .
\end{eqnarray}

\subsection{Motion with an inverse square dependence}

In the case of  an inverse square   density profile
for the ISM,
as given by equation (\ref{profsquare}),
the differential equation
that models the momentum conservation
is
\begin{eqnarray}
\Bigg( \frac{1}{3}\,\rho_{{0}}{r_{{0}}}^{3}+{\frac {\rho_{{0}}{z_{{0}}}^{2}r
 \left( t \right) }{ \left( \cos \left( \theta \right)  \right) ^{2}}}
-2\,{\frac {\rho_{{0}}{z_{{0}}}^{3}\ln  \left( r \left( t \right) \cos
 \left( \theta \right) +z_{{0}} \right) }{ \left( \cos \left( \theta
 \right)  \right) ^{3}}}-{\frac {\rho_{{0}}{z_{{0}}}^{4}}{ \left( \cos
 \left( \theta \right)  \right) ^{3} \left( r \left( t \right) \cos
 \left( \theta \right) +z_{{0}} \right) }}
\nonumber \\
-{\frac {\rho_{{0}}{z_{{0}}}
^{2}r_{{0}}}{ \left( \cos \left( \theta \right)  \right) ^{2}}}+2\,{
\frac {\rho_{{0}}{z_{{0}}}^{3}\ln  \left( r_{{0}}\cos \left( \theta
 \right) +z_{{0}} \right) }{ \left( \cos \left( \theta \right) 
 \right) ^{3}}}+{\frac {\rho_{{0}}{z_{{0}}}^{4}}{ \left( \cos \left( 
\theta \right)  \right) ^{3} \left( r_{{0}}\cos \left( \theta \right) 
+z_{{0}} \right) }} \Bigg ) {\frac {\rm d}{{\rm d}t}}r \left( t
 \right) -\frac{1}{3}\,\rho_{{0}}{r_{{0}}}^{3}v_{{0}}=0
\quad .
\end{eqnarray}
There is not an analytical solution for this differential equation.

\subsection{Motion with a Navarro--Frenk--White profile}

\begin{figure*}
\begin{center}
\includegraphics[width=5cm,angle=-90]{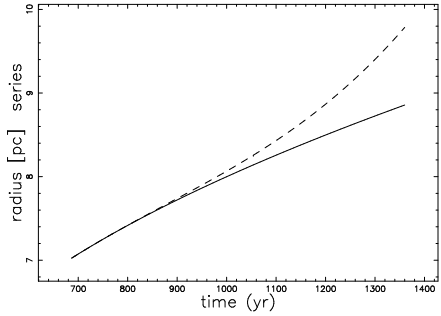}
\end {center}
\caption
{
Numerical solution (full line) and  series solution (dashed line)
The parameters are 
$v_0\,=3700$ km s$^{-1}$,
$r_0\,=7 \,pc$,
$b\,=2.8 \,pc$,
$t=1360  \,yr$ and
$t_0=680 \,yr$.
}
\label{localb_nfw_series}
    \end{figure*}

In the case of  a Navarro--Frenk--White density profile
for the ISM,
as given by equation (\ref{profilenfwz}),
the differential equation
that models momentum conservation
is 
\begin{eqnarray}
 \Bigg( \frac{1}{3}\,\rho_{{0}}{r_{{0}}}^{3}+{\frac {r_{{0}}\rho_{{0}} \left( 
 \left( b+r \left( t \right) \cos \left( \theta \right)  \right) \ln 
 \left( b+r \left( t \right) \cos \left( \theta \right)  \right) +b
 \right)  \left( b+r_{{0}} \right) ^{2}}{ \left( \cos \left( \theta
 \right)  \right) ^{3} \left( b+r \left( t \right) \cos \left( \theta
 \right)  \right) }}
\nonumber \\
-{\frac {r_{{0}}\rho_{{0}} \left(  \left( b+r_{{0}
}\cos \left( \theta \right)  \right) \ln  \left( b+r_{{0}}\cos \left( 
\theta \right)  \right) +b \right)  \left( b+r_{{0}} \right) ^{2}}{
 \left( \cos \left( \theta \right)  \right) ^{3} \left( b+r_{{0}}\cos
 \left( \theta \right)  \right) }} \Bigg) {\frac {\rm d}{{\rm d}t}}r
 \left( t \right) =\frac{1}{3}\,\rho_{{0}}{r_{{0}}}^{3}v_{{0}}
\quad .
\end{eqnarray}
A {\it first} approximated solution  
of this differential equation can be
given as a series  of  order 4 
\begin{eqnarray}
r(t;t_0,r_0,v_0,b) =
r_{{0}}+v_{{0}} \left( t-{\it t_0} \right) 
-\frac{3}{2}\,{\frac { \left( b+r_{{0
}} \right) ^{2}{v_{{0}}}^{2} \left( t-{\it t_0} \right) ^{2}}{r_{{0}}
\cos \left( \theta \right)  \left( b+r_{{0}}\cos \left( \theta
 \right)  \right) ^{2}}}
\nonumber  \\
+\frac{1}{2}\,{\frac { \left( b+r_{{0}} \right) ^{2}{v
_{{0}}}^{3} \left(  \left( \cos \left( \theta \right)  \right) ^{3}{r_
{{0}}}^{2}-\cos \left( \theta \right) {b}^{2}+9\,{r_{{0}}}^{2}+18\,br_
{{0}}+9\,{b}^{2} \right)  \left( t-{\it t_0} \right) ^{3}}{{r_{{0}}}^{2
} \left( \cos \left( \theta \right)  \right) ^{2} \left( b+r_{{0}}\cos
 \left( \theta \right)  \right) ^{4}}}
\quad .
\end{eqnarray}
Figure \ref{localb_nfw_series} reports a comparison 
between numerical and  series solution.
% figure   localb_nfw_series
% figure   localb_nfw_series

To find a {\it second} approximate solution 
for 
this differential equation of the first order in $r$,
we separate the
variables  and we integrate.
The  following non-linear equation is obtained
\begin{equation}
\frac{N}{D} = t-t_0
\label{eqn_nl_nfw}
\quad ,
\end{equation}
where 
\begin{eqnarray}
N =-6\, \left( b+r_{{0}} \right) ^{2} \left( b+r_{{0}}\cos \left( \theta
 \right)  \right)  \left( 1/2\,r\cos \left( \theta \right) 
+b \right) 
\ln  \left( b+r_{{0}}\cos \left( \theta \right)  \right) 
\nonumber  \\
+6\, \left( b
+r_{{0}} \right) ^{2} \left( b+r_{{0}}\cos \left( \theta \right) 
 \right)  \left( 1/2\,r\cos \left( \theta \right) +b \right) \ln 
 \left( b+r\cos \left( \theta \right)  \right) 
\nonumber \\ 
-6\,\cos \left( \theta
 \right)  \left( r-r_{{0}} \right)  \left( -1/6\,{r_{{0}}}^{3} \left( 
\cos \left( \theta \right)  \right) ^{4}-1/6\,b{r_{{0}}}^{2} \left( 
\cos \left( \theta \right)  \right) ^{3}+1/2\,r_{{0}} \left( b+r_{{0}}
 \right) ^{2}\cos \left( \theta \right) +b \left( b+r_{{0}} \right) ^{
2} \right) 
\quad ,
\end{eqnarray}
and
\begin{eqnarray}
D = v_{{0}}{r_{{0}}}^{2} \left( \cos \left( \theta \right)  \right) ^{4}
 \left( b+r_{{0}}\cos \left( \theta \right)  \right) 
\quad .
\end{eqnarray}

In this case, it is not possible to find an analytical  
solution for the radius, $r$,
as a function of  time.
Therefore, we apply 
the Pad\'e rational polynomial, see
\cite{Pade1892,Wynn1966,Baker1975,NIST2010}.  
We choose an approximation of degree 2 in the numerator
and degree 1 in the denominator about the point $r=r_0$ 
to the  left-hand  side of 
equation~(\ref{eqn_nl_nfw}).
The resulting equation of second degree is 
\begin{equation}
\frac{NN}{DD} = t-t_0
\quad ,
\end{equation}
where 
\begin{eqnarray}
NN =
- \Big  ( r_{{0}}-r \Big  )  \Big  ( 4\,   ( \cos   ( \theta
   )    ) ^{3}{r_{{0}}}^{3}+2\,   ( \cos   ( \theta
   )    ) ^{3}{r_{{0}}}^{2}r+12\,   ( \cos   ( \theta
   )    ) ^{2}{r_{{0}}}^{2}b+8\,r_{{0}}\cos   ( \theta
   ) {b}^{2}
\nonumber \\
-2\,\cos   ( \theta   ) {b}^{2}r-9\,{r_{{0}}}^{
3}-18\,b{r_{{0}}}^{2}+9\,{r_{{0}}}^{2}r-9\,r_{{0}}{b}^{2}+18\,r_{{0}}b
r+9\,{b}^{2}r  \Big ) 
\quad ,
\end{eqnarray}
and
\begin{eqnarray}
DD = 
2\,\cos \left( \theta \right) v_{{0}} \left( b+r_{{0}}\cos \left( 
\theta \right)  \right)  \left( 2\,\cos \left( \theta \right) {r_{{0}}
}^{2}+r_{{0}}r\cos \left( \theta \right) +4\,br_{{0}}-br \right) 
\quad .
\end{eqnarray}
The resulting Pad\'e  approximant  for the trajectory ,the radius $r_{2,1}$,
 is
the  {\it second} approximated solution
\begin{eqnarray}
r_{2,1}(t;t_0,r_0,v_0,b)=
\frac{NNN}{DDD}
\label{rmotionnfw} 
\quad .
\end{eqnarray}
where 
\begin{eqnarray}
NNN = 
 \left( \cos \left( \theta \right)  \right) ^{3}t{r_{{0}}}^{2}v_{{0}}-
 \left( \cos \left( \theta \right)  \right) ^{3}{r_{{0}}}^{2}t_{{0}}v_
{{0}}- \left( \cos \left( \theta \right)  \right) ^{3}{r_{{0}}}^{3}-6
\, \left( \cos \left( \theta \right)  \right) ^{2}{r_{{0}}}^{2}b-\cos
 \left( \theta \right) {b}^{2}tv_{{0}}
\nonumber \\
+\cos \left( \theta \right) {b}^
{2}t_{{0}}v_{{0}}-5\,r_{{0}}\cos \left( \theta \right) {b}^{2}+9\,r_{{0
}}{b}^{2}+18\,{r_{{0}}}^{2}b+9\,{r_{{0}}}^{3}+\sqrt {A}
\quad ,
\end{eqnarray}
and
\begin{eqnarray}
DDD =
2\, \left( \cos \left( \theta \right)  \right) ^{3}{r_{{0}}}^{2}-2\,{b
}^{2}\cos \left( \theta \right) +9\,{b}^{2}+18\,br_{{0}}+9\,{r_{{0}}}^
{2}
\quad ,
\end{eqnarray}
and
\begin{eqnarray}
A=
  ( b+r_{{0}}\cos   ( \theta   )    ) ^{2}\cos   ( 
\theta   )  
  \Big (    ( 3\,r_{{0}}+v_{{0}}   ( t-t_{{0}}
   )    ) ^{2}{r_{{0}}}^{2}   ( \cos   ( \theta   ) 
   ) ^{3}
\nonumber \\
-2\,   ( 3\,r_{{0}}+v_{{0}}   ( t-t_{{0}}   ) 
   )    ( -3\,r_{{0}}+v_{{0}}   ( t-t_{{0}}   ) 
   ) r_{{0}}b   ( \cos   ( \theta   )    ) ^{2}+
   ( -3\,r_{{0}}+v_{{0}}   ( t-t_{{0}}   )    ) ^{2}{b}^
{2}\cos   ( \theta   ) 
\nonumber \\
+54\,r_{{0}}v_{{0}}   ( b+r_{{0}}
   ) ^{2}   ( t-t_{{0}}   )  \Big  ) 
\quad  .
\end{eqnarray}

\begin{figure*}
\begin{center}
\includegraphics[width=5cm,angle=-90]{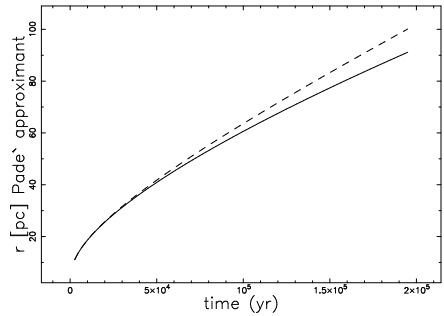}
\end {center}
\caption
{
Numerical solution (full line) and Pad\'e  approximant  (dashed line)
The parameters are 
$v_0\,=3700$ km s$^{-1}$,
$r_0\,=7 \,pc$,
$b\,=2.8 \,pc$,
$t=194285  \,yr$ and
$t_0=680 \,yr$.
}
\label{localb_pade_nfw}
    \end{figure*}

Figure \ref{localb_pade_nfw} reports a comparison 
between the numerical and the series solution.
% figure   localb_pade_nfw
% figure   localb_pade_nfw

The two approximations that we have used here cover 
the range in time after which the  
percent  error is $\approx 10 \%$: 1360 $yr$ 
for the Taylor series and 194285 $yr$ for  the Pad\'e  approximant.
We conclude that in our case the Pad\'e  approximant has a wider radius
of convergence in respect to the  Taylor series.

\section{Astrophysical results}
\label{section_astro}

The adopted astrophysical units  are pc
for length and
$yr$ for time; while
the initial velocity $v_{{0}}$ is
expressed in
pc yr$^{-1}$.
The astronomical velocities are evaluated in
km s$^{-1}$
and  therefore
$v_{{0}} =1.02\times10^{-6} v_{{1}}$
where  $v_{{1}}$ is the initial
velocity expressed in
km s$^{-1}$.

\subsection{How to start }

The  starting equations for the evolution
of the SB \cite{Dyson1997,mccrayapj87}
are defined by the following parameters:
$N^*$, which is
the number of SN explosions in  $5.0 \cdot 10^7$ \mbox{yr};
$Z_{\mathrm{OB}}$, which is
the distance of the OB associations from the galactic plane;
$E_{51}$, which is
the  energy in  $10^{51}$ \mbox{erg} and is usually chosen equal to one;
$v_0$, which is
the initial velocity, which is fixed by the bursting phase,
$t_0$;
the initial time in $yr$,  which is equal to the bursting time; 
and $t$, which is  the proper time  of the SB.
The  radius of the SB 
is
\begin{equation}
R =111.56\,(\frac{E_{51}t_7^3 N^*}{n_0})^{\frac {1} {5}}
\,\mathrm{pc},
\label{raggioburst}
\end{equation}
and its velocity 
\begin{equation}
V= 6.567\,{\frac {1}{{{\it t_7}}^{2/5}}\sqrt [5]{{\frac {E_{{51}}
{\it N^*}}{n_{{0}}}}}} \,\mathrm{ \frac{km}{s}} 
\quad .
\end{equation}
In the following, we will  assume that 
the bursting phase  ends at $t=t_{7,0}$   (the bursting time is expressed in
units of $10^7$ yr) 
when  $N_{SN}$ SN are exploded 
\begin{equation}
N_{SN} = N^* \frac{t_{7,0} \cdot  10^7} {5 \cdot 10^7}
\quad .
\end{equation}   
The two following  inverted formula allow us to derive  
the parameters of the initial conditions for the SB
in terms of  $r_0$ expressed in pc  and $v_0$ expressed 
in $km\,s^{-1}$ 
\begin{equation}
t_{7,0}= 0.05878\,{\frac {r_{{0}}}{v_{{0}}}} 
\quad ,
\end{equation}
and
\begin{equation}
N^*= 2.8289\,10^{-7}\,{\frac {{r_{{0}}}^{2}n_{{0}}{v_{{0}}}^{3}}{E_{{51
}}}}
\quad  .
\end{equation}

\subsection{The astronomical data}

The LB has been recently observed in the X-ray in the 
$0.1-1.2\,kev$ region by \cite{Liu2017}, whose
Figure 7  
reports six  configurations   of the LB along great-circle cuts through
the Galactic pole and the Galactic plane.
As a target of the simulation, we have chosen the cut
characterized  by galactic longitude, $l$, between
120 $^\circ$ and  300 $^\circ$. 
An observational
percentage reliability, $\epsilon_{\mathrm {obs}}$,
is  introduced over the whole range
of the polar   angle  $\theta$,
\begin{equation}
\epsilon_{\mathrm {obs}}  =100(1-\frac{\sum_j |r_{\mathrm {obs}}-r_{\mathrm{num}}|_j}{\sum_j
{r_{\mathrm {obs}}}_{,j}})
\quad,
\label{efficiencymany}
\end{equation}
where
$r_{\mathrm{num}}$ is the theoretical radius of the local LB,
$r_{\mathrm{obs}}$ is the observed    radius of the local LB, 
and
the  index $j$  varies  from 1 to the number of
available observations.
The observational
percentage of reliability  allows us to fix the theoretical parameters.

\subsection{The results}

\begin{figure*}
\begin{center}
\includegraphics[width=5cm,angle=-90]{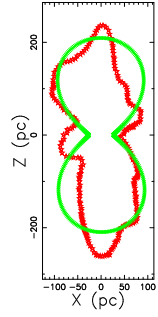}
\end {center}
\caption
{
Geometrical section of the LB 
in the $x-z$ plane with an exponential profile
(green points)
and observed profile
(red stars).
The parameters are 
$v_0\,=3700$ km s$^{-1}$,
$r_0\,=7 \,pc$,
$b  \,=3.5 \,pc$,
$t=8\,10^4\,yr$ and
$t_0=80 \,yr$.
The observational
percentage reliability is $\epsilon_{\mathrm {obs}}=81.93 \%$
and $N_{SN}$=15.61.
}
\label{localb_theo_obs_exp}
    \end{figure*}

\begin{figure*}
\begin{center}
\includegraphics[width=5cm,angle=-90]{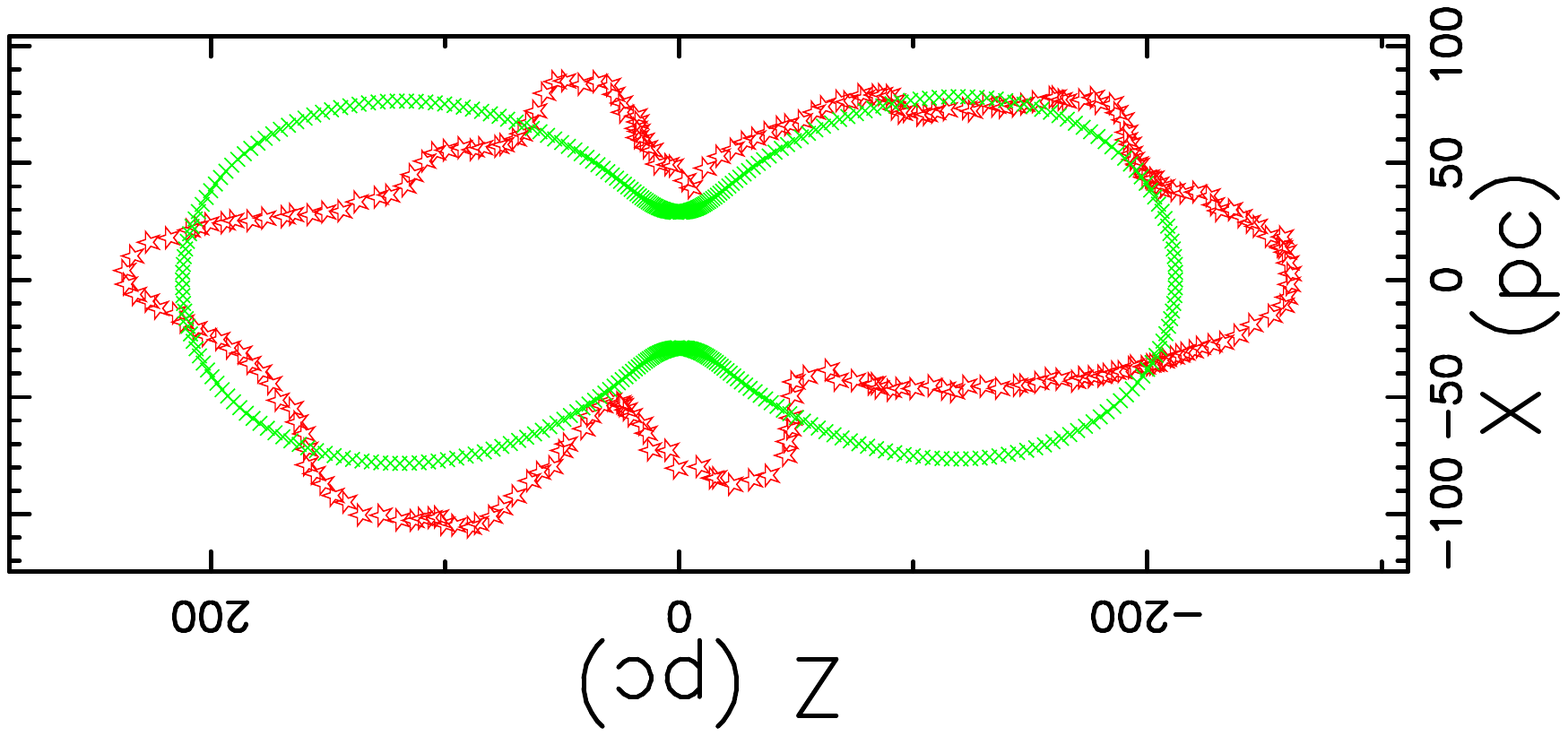}
\end {center}
\caption
{
Geometrical section of the LB 
in the $x-z$ plane with a Gaussian profile
(green points)
and observed profile
(red stars).
The parameters are 
$v_0\,=3700$ km s$^{-1}$,
$r_0\,=7 \,pc$,
$b  \,=5.83 \,pc$,
$t=1.35\,10^5\,yr$ and
$t_0=1.35\,10^2 \,yr$.
The observational
percentage reliability is $\epsilon_{\mathrm {obs}}=82.04 \%$
and $N_{SN}$=15.61.
}
\label{localb_theo_obs_gauss}
    \end{figure*}

\begin{figure*}
\begin{center}
\includegraphics[width=5cm,angle=-90]{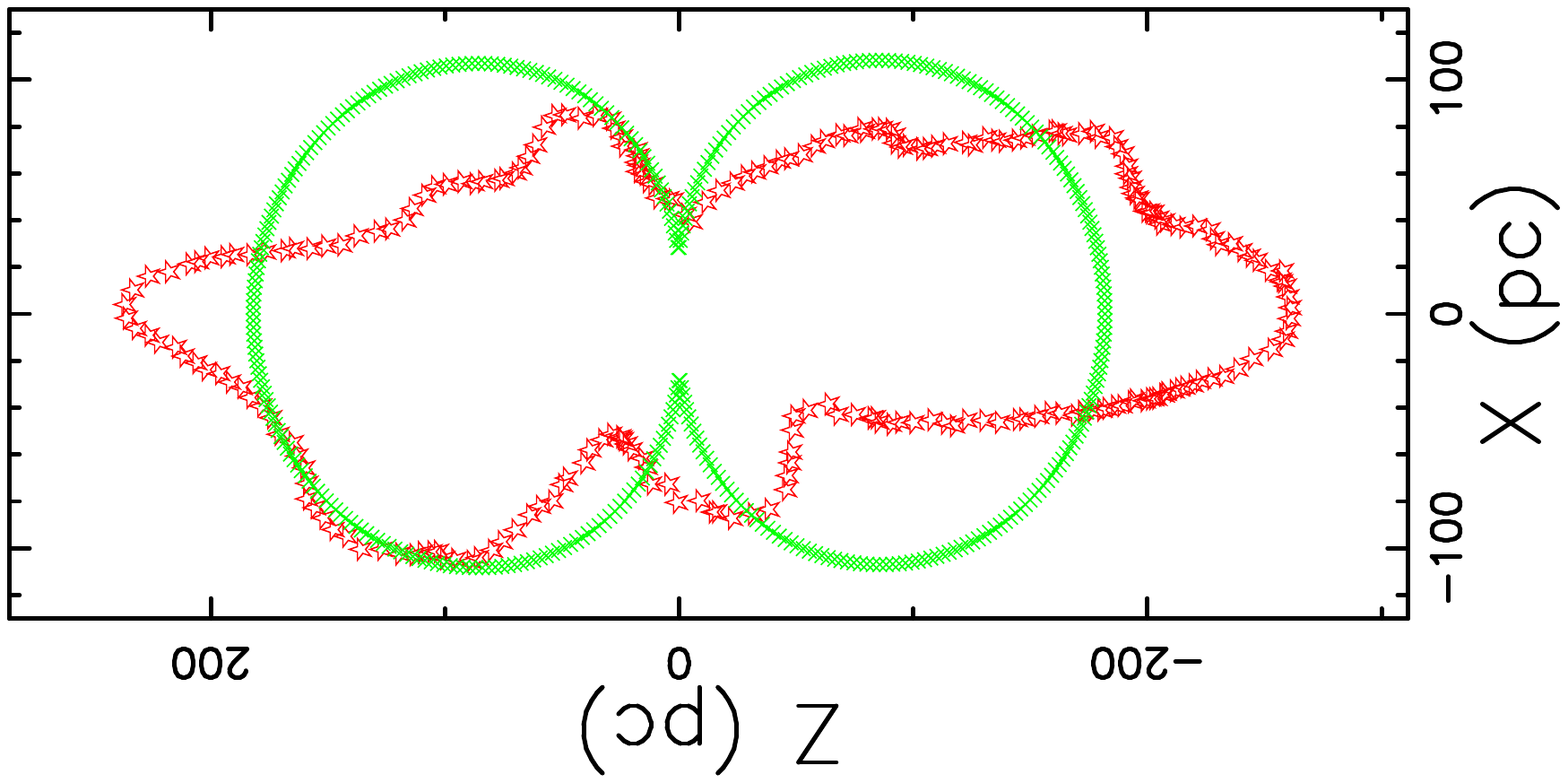}
\end {center}
\caption
{
Geometrical section of the LB 
in the $x-z$ plane with an exponential profile
(green points)
and observed profile
(red stars).
The parameters are 
$v_0\,=3700$ km s$^{-1}$,
$r_0\,=7 \,pc$,
$z_0\,=7 \,pc$,
$t=8\,10^4\,yr$ and
$t_0=80 \,yr$.
The observational
percentage reliability is $\epsilon_{\mathrm {obs}}=78.02 \%$
and $N_{SN}$=15.61.
}
\label{localb_theo_obs_square}
    \end{figure*}

\begin{figure*}
\begin{center}
\includegraphics[width=5cm,angle=-90]{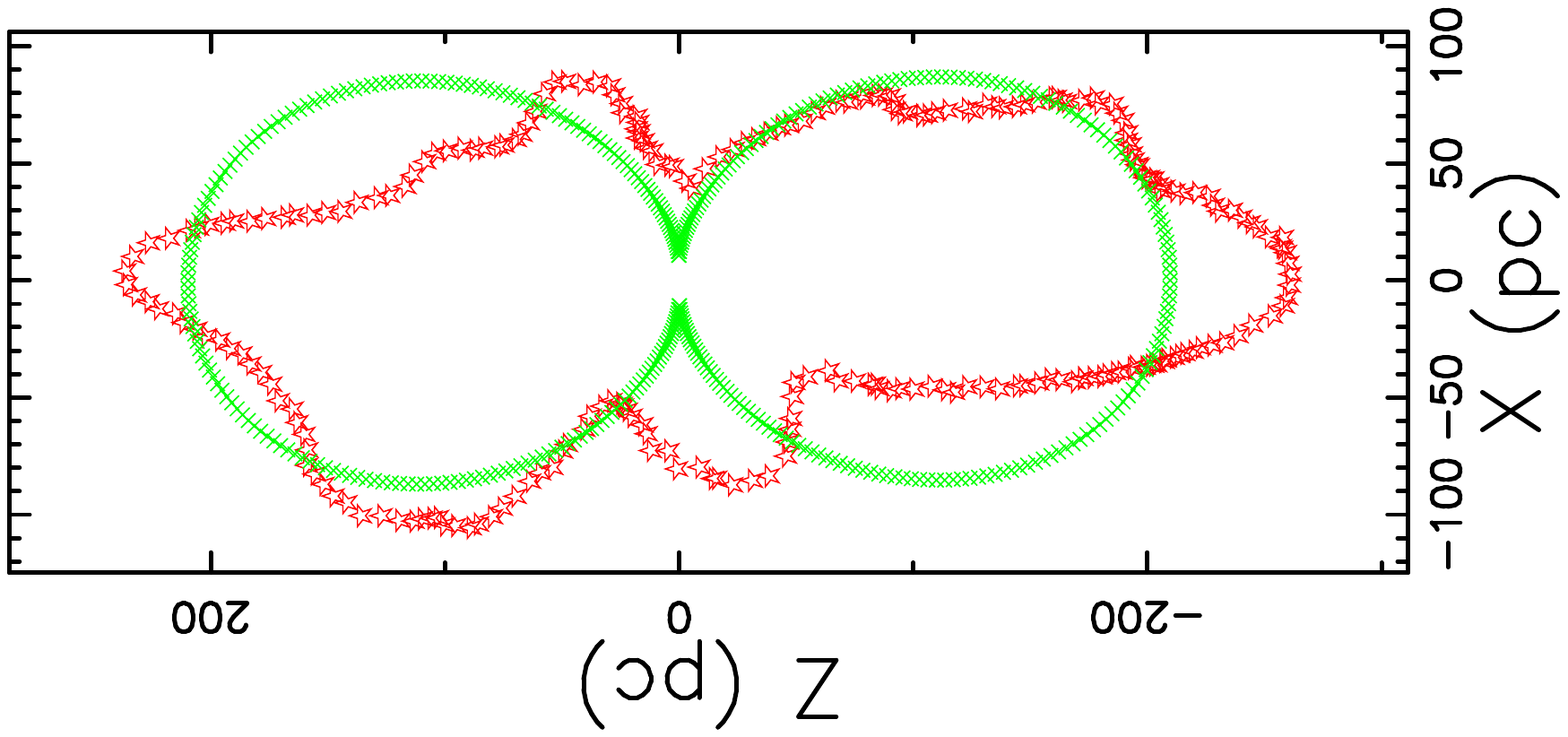}
\end {center}
\caption
{
Geometrical section of the LB 
in the $x-z$ plane with a  NFW  profile
(green points)
and observed profile
(red stars).
The parameters are 
$v_0\,=3700$ km s$^{-1}$,
$r_0\,=7 \,pc$,
$b\,=2.8 \,pc$,
$t=6.8\,10^5\,yr$ and
$t_0=680 \,yr$.
The observational
percentage reliability is $\epsilon_{\mathrm {obs}}=82.69 \%$
and $N_{SN}$=15.61.
}
\label{localb_theo_obs_nfw}
    \end{figure*}

The  numerical solution is reported 
as a cut   in 
the $x-z$ plane:
see 
Figure \ref{localb_theo_obs_exp} 
for an exponential density profile
as given by equation (\ref{profexponentialr});
see Figure \ref{localb_theo_obs_gauss} 
for a Gaussian density profile
as given by equation (\ref{profgaussian});
see Figure \ref{localb_theo_obs_square} 
for an inverse square density profile
as given by equation (\ref{profsquare});
and
see Figure \ref{localb_theo_obs_nfw} 
for a  NFW  density profile
as given by equation (\ref{profilenfwz}).
% figure   localb_theo_obs_exp
% figure   localb_theo_obs_exp

% figure   localb_theo_obs_gauss
% figure   localb_theo_obs_gauss

% figure   localb_theo_obs_square
% figure   localb_theo_obs_square

% figure   localb_theo_obs_nfw
% figure   localb_theo_obs_nfw

\begin{figure*}
\begin{center}
\includegraphics[width=7cm]{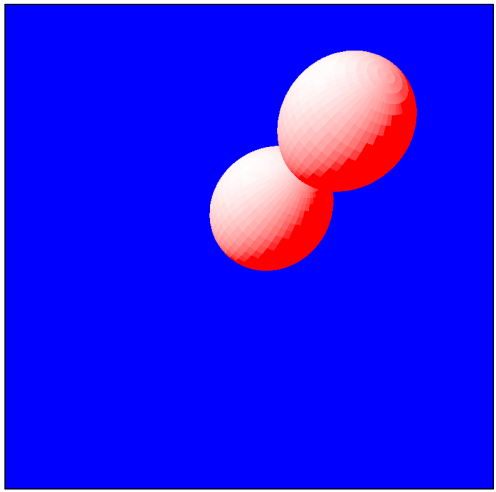}
\end {center}
\caption
{
3D surface  of the LB 
with parameters as in Figure  \ref{localb_theo_obs_nfw}, 
NFW  profile.
The three Euler angles are $\Theta=41$, $\Phi=-41$ and
$ \Psi=41 $.
}
\label{localb_3d_nfw}
    \end{figure*}

The theory of the asymmetrical expansion 
already developed is independent  
of  the  
azimuthal angle $\varphi$ and therefore 
the 3D advancing surface of a LB   can be obtained  
by rotating
a cut in $x-z$ plane,
see Figure \ref{localb_3d_nfw}.

% figure   localb_3d_nfw
% figure   localb_3d_nfw

\subsection{$^{60}$Fe-signature} 

\begin{figure*}
\begin{center}
\includegraphics[width=5cm]{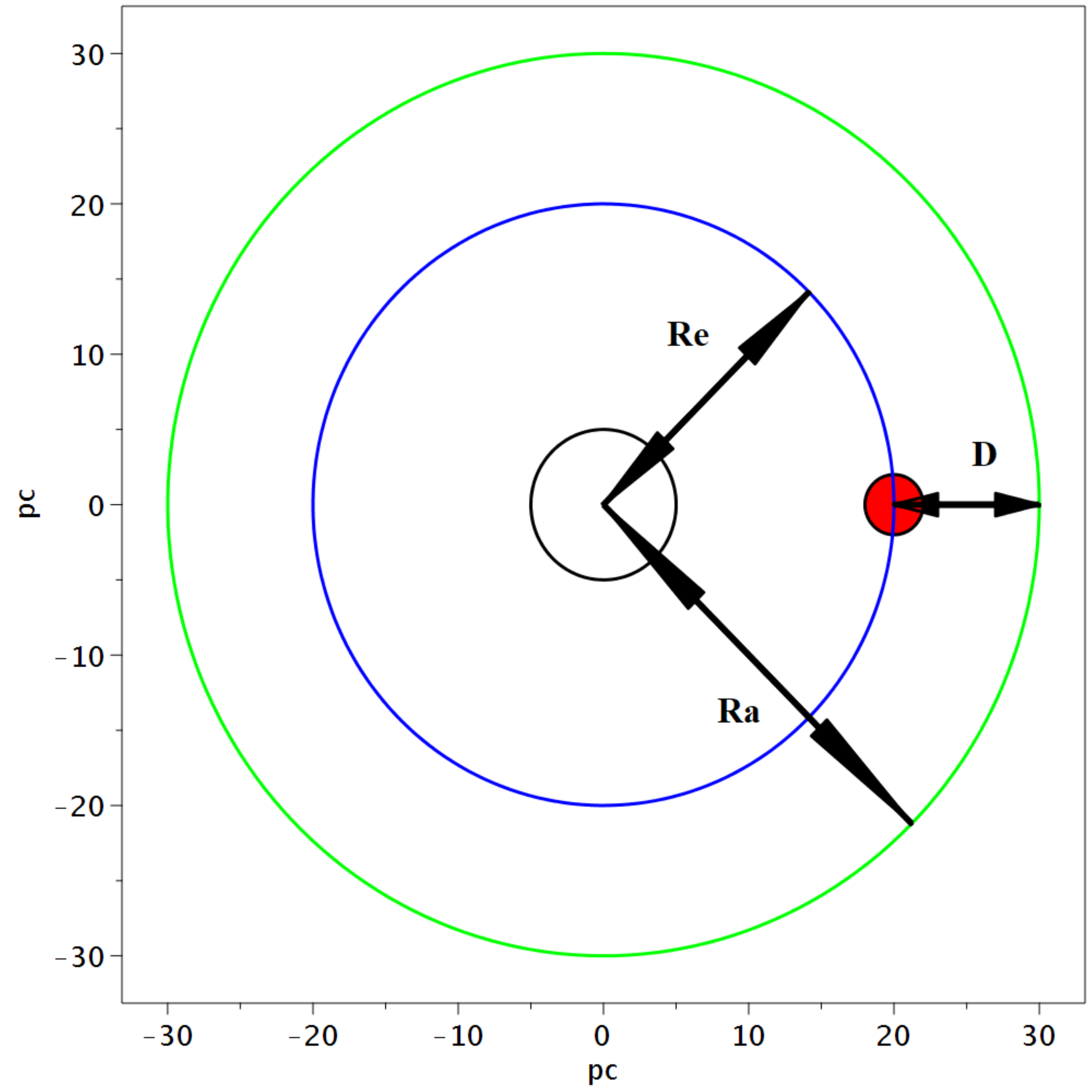}
\end {center}
\caption
{
A sketch of the LB sun encounter.
The black circle  is the initial radius,
the blue  circle  is the radius when the Sun is reached
and green circle  is the actual radius. 
}
\label{localb_sun}
    \end{figure*}

Some radioisotopes on Earth, such as $^{60}$Fe 
(half life of $1.5\,10^6 yr$ \cite{Rugel2009}), 
were measured 
in a deep-sea ferromanganese crust:
the concentration  of $^{60}$Fe increased 2.8 $Myr$ ago,
see \cite{Knie2004}.
These measurements have  triggered 
some  simulations that can explain the LB
in the framework of  SN explosions \cite{Feige2017a,Feige2017b,Feige2018}.
The encounter  between the advancing shell of the LB and the Sun
is here simulated in 2D 
assuming a constant density, see equation (\ref{rtconstant}).
The  following  distances are involved: 
\begin{enumerate}
\item
$r_0$ the initial radius of the LB,
\item 
$r_e$ the radius of the LB when encounters the LB,
\item
$r_a$ the actual radius of the LB,   
\item
$D$ the distance between the sun and the LB, $D=r_a-r_e$,  
\end{enumerate}
and they  are reported in  Figure \ref{localb_sun}.
The times of the 2D simulation are  
\begin{enumerate}
\item
$t_0$ the time at which the radius of the LB is $r_0$,
\item 
$t_{^{60}Fe}$ the time at which $^{60}$Fe  was deposited on the 
         Earth,
\item
$t_a$ the actual time of the LB,
\item 
$t_e$ the time of the encounter between  LB and Sun, 
$t_e=t_a -t_{^{60}Fe}$.

\end{enumerate}

% figure   localb_sun
\begin{figure*}
\begin{center}
\includegraphics[width=5cm,angle=-90]{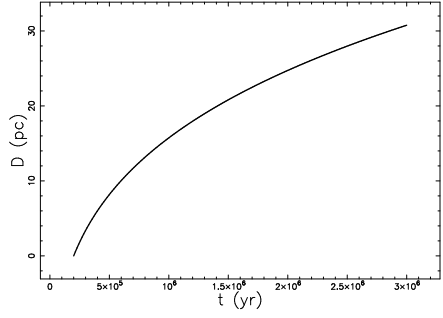}
\end {center}
\caption
{
Distance between Sun and the LB as function of time.
The parameters of the LB are 
$v_0\,=10000$ km s$^{-1}$,
$r_0\,=5 \,pc$, 
$t_{max} = 3  \, 10^6\,yr$,
$t_{^{60}Fe}= 2.8\, 10^6\,yr$
and  
$t_0=100 \,yr$.
}
\label{localb_fe60}
    \end{figure*}

% figure   localb_sun
The distance LB-Sun, $D$, is reported 
in Figure \ref{localb_fe60} as function
of the elapsed time.
% figure   localb_fe60
% figure   localb_fe60

\subsection{Collective effects}

\begin{figure*}
\begin{center}
\includegraphics[width=5cm]{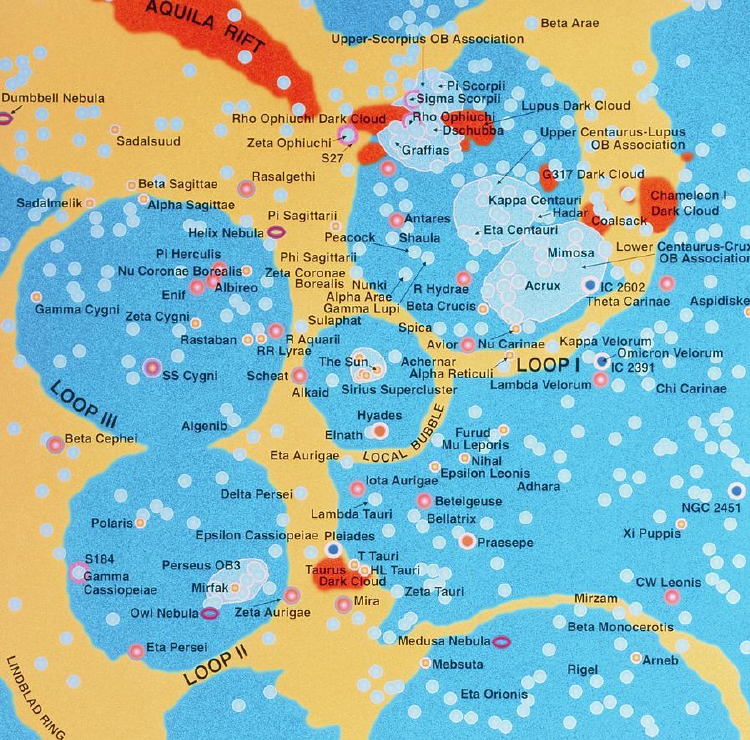}
\end {center}
\caption
{
Map of the galactic environment of the sun
with side of $\approx 528.12$ pc. 
}
\label{realbubbles}
    \end{figure*}

\begin{figure*}
\begin{center}
\includegraphics[width=5cm,angle=-90]{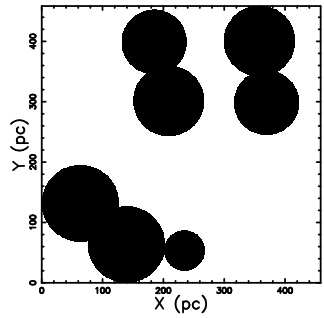}
\end {center}
\caption
{
Network of seven bubbles,
$v_0\,=5000$ km s$^{-1}$,
$r_0\,=7 \,pc$, 
$t_{max} = 3\, 10^6\,yr$, 
$t_0=100 \,yr$
and $side=400\,pc$.
}
\label{localb_many}
    \end{figure*}

The LB   is a part of  other bubbles which
show
a Swiss–cheese  structure, see Figure \ref{realbubbles}.
% figure   realbubbles
% figure   realbubbles
We simulate this network 
with the multiple explosion of $N$ bubbles in 2D 
assuming a constant density, see equation (\ref{rtconstant}).
We choose $N=7$ and the 
time is allowed to vary in a random way in the
interval $(t_0,t_{max})$,
the position of the  explosion 
on  the two Cartesian axis 
is  randomly generated in the interval $(0,side)$,
see Figure \ref{localb_many}.
  
% figure   localb_many
% figure   localb_many

%siamoqui
\section{Conclusions}

Two factors allow  the comparison of different 
models  which  simulate the LB: 
the observational
percentage reliability, see  equation (\ref{efficiencymany});
and acceptable observational cuts of the LB, see \cite{Liu2017}.
The best result is obtained adopting 
the NFW  profile
with a percentage reliability of $\epsilon_{\mathrm {obs}}=82.69 \%$.
Similar results are obtained  in the framework  of the 
magnetic field model, see Figure 2 in \cite{Alves2018}.
The $^{60}$Fe-signature is compatible with the model that we have
developed  here and Figure \ref{localb_fe60}
reports the distance between the Sun and the LB.
A simulation of  the  exploding bubbles is reported
in Figure \ref{localb_many}.
A more precise simulation of the exploding bubbles
can be done when more detailed observations 
of the network, such as that reported in Figure 
\ref{realbubbles}, are available.

\section*{Acknowledgments}

Credit for Figure
\ref{realbubbles}
 is  given to  
the  University of Bologna,
see \url{https://www.sslmit.unibo.it/zat/images/cartography/M-Way_2.htm}.

%\bibliography{biblio}

\begin{thebibliography}{10}
\expandafter\ifx\csname url\endcsname\relax
  \def\url#1{{\tt #1}}\fi
\expandafter\ifx\csname urlprefix\endcsname\relax\def\urlprefix{URL }\fi
\providecommand{\eprint}[2][]{\url{#2}}
% Bibliography created with iopart-num.bst, v1.0

\bibitem{Arnaud1984}
{Arnaud} M, {Rothenflug} R and {Rocchia} R 1984 {The local hot bubble from
  X-ray spectroscopic measurements.} {\em Physica Scripta Volume T\/} {\bf 7},
  48

\bibitem{Slavin2016}
Slavin J~D 2016 in P~Alsabti Athem Wand~Murdin, ed, {\em Handbook of
  Supernovae\/} (Cham: Springer International Publishing) pp 1--13

\bibitem{Liu2017}
{Liu} W, {Chiao} M, {Collier} M~R and et~al 2017 {The Structure of the Local
  Hot Bubble} {\em The Astrophysical Journal\/} {\bf 834}(1) 33
  (\textit{Preprint} \eprint{1611.05133})

\bibitem{Gontcharov2019}
{Gontcharov} G~A and {Mosenkov} A~V 2019 {Interstellar polarization and
  extinction in the Local Bubble and the Gould Belt} {\em \mnras\/} {\bf
  483}(1), 299 (\textit{Preprint} \eprint{1811.01411})

\bibitem{Medan2019}
{Medan} I and {Andersson} B~G 2019 {Magnetic Field Strengths and Variations in
  Grain Alignment in the Local Bubble Wall} {\em \apj\/} {\bf 873}(1) 87
  (\textit{Preprint} \eprint{1901.07692})

\bibitem{Knie2004}
{Knie} K, {Korschinek} G, {Faestermann} T, {Dorfi} E~A, {Rugel} G and {Wallner}
  A 2004 {$^{60}$Fe Anomaly in a Deep-Sea Manganese Crust and Implications for
  a Nearby Supernova Source} {\em \prl\/} {\bf 93}(17) 171103

\bibitem{Smith1998}
{Smith} R~K and {Cox} D~P 1998 in D~{Breitschwerdt}, M~J {Freyberg} and
  J~{Truemper}, eds, {\em Lecture Notes in Physics, vol.506, The Local Bubble
  and Beyond. Lyman-Spitzer Colloquium, Proceedings of the IAU Colluquium No.
  166 held in Garching, Germany\/} vol 506 (Berlin: Springer-Verlag) pp
  133--136

\bibitem{Smith2001}
{Smith} R~K and {Cox} D~P 2001 {Multiple Supernova Remnant Models of the Local
  Bubble and the Soft X-Ray Background} {\em \apjs\/} {\bf 134}(2), 283

\bibitem{Welsh2002}
{Welsh} B~Y, {Sallmen} S and {Lallement} R 2002 {New results from FUSE: a
  paradigm for testing models of the Local Hot Bubble} in {\em American
  Astronomical Society Meeting Abstracts \#200\/} vol 200 of {\em American
  Astronomical Society Meeting Abstracts\/} pp 767--778

\bibitem{Breitschwerdt2009}
{Breitschwerdt} D, {de Avillez} M~A and {Baumgartner} V 2009 {Modeling the
  Local Warm/Hot Bubble} in R~K {Smith}, S~L {Snowden} and K~D {Kuntz}, eds,
  {\em American Institute of Physics Conference Series\/} vol 1156 of {\em
  American Institute of Physics Conference Series\/} pp 271--279
  (\textit{Preprint} \eprint{0812.0505})

\bibitem{Deavillez2012}
{de Avillez} M~A and {Breitschwerdt} D 2012 {Non-equilibrium ionization
  modeling of the Local Bubble. I. Tracing Civ, Nv, and Ovi ions} {\em \aap\/}
  {\bf 539} L1

\bibitem{Feige2018}
{Schulreich} M, {Breitschwerdt} D, {Feige} J and {Dettbarn} C 2018 {A Way Out
  of the Bubble Trouble?{\textemdash}Upon Reconstructing the Origin of the
  Local Bubble and Loop I via Radioisotopic Signatures on Earth} {\em
  Galaxies\/} {\bf 6}(1), 26 (\textit{Preprint} \eprint{1802.09275})

\bibitem{Alves2018}
{Alves} M~I~R, {Boulanger} F, {Ferri{\`e}re} K and {Montier} L 2018 {The Local
  Bubble: a magnetic veil to our Galaxy} {\em \aap\/} {\bf 611} L5
  (\textit{Preprint} \eprint{1803.05251})

\bibitem{Navarro1996}
{Navarro} J~F, {Frenk} C~S and {White} S~D~M 1996 {The Structure of Cold Dark
  Matter Halos} {\em \apj\/} {\bf 462}, 563 (\textit{Preprint}
  \eprint{astro-ph/9508025})

\bibitem{Dyson1997}
{{Dyson}, J~E and {Williams}, D~A} 1997 {\em {The physics of the interstellar
  medium}\/} (Bristol: Institute of Physics Publishing)

\bibitem{Padmanabhan_II_2001}
{Padmanabhan} P {2001} {\em {Theoretical astrophysics. Vol. II: Stars and
  Stellar Systems}\/} ({Cambridge, UK}: {Cambridge University Press})

\bibitem{Pade1892}
{Pad{\'e}} H 1892 Sur la repr{\'e}sentation approch{\'e}e d'une fonction par
  des fractions rationnelles {\em Ann. Sci. Ecole Norm. Sup.\/} {\bf 9}, 193

\bibitem{Wynn1966}
Wynn P 1966 Upon systems of recursions which obtain among the quotients of the
  pad{\'e} table {\em Numerische Mathematik\/} {\bf 8}(3), 264

\bibitem{Baker1975}
{Baker} G 1975 {\em Essentials of Pad{\'e} approximants\/} (New York: Academic
  Press)

\bibitem{NIST2010}
Olver F~W~J~e, Lozier D~W~e, Boisvert R~F~e and Clark C~W~e 2010 {\em {NIST
  handbook of mathematical functions.}\/} (Cambridge: {Cambridge University
  Press. })

\bibitem{mccrayapj87}
{McCray} R and {Kafatos} M 1987 {Supershells and propagating star formation}
  {\em \apj\/} {\bf 317}, 190

\bibitem{Rugel2009}
{Rugel} G, {Faestermann} T, {Knie} K, {Korschinek} G, {Poutivtsev} M,
  {Schumann} D, {Kivel} N, {G{\"u}nther-Leopold} I, {Weinreich} R and
  {Wohlmuther} M 2009 {New Measurement of the Fe60 Half-Life} {\em \prl\/} {\bf
  103}(7) 072502

\bibitem{Feige2017a}
{Feige} J, {Breitschwerdt} D, {Wallner} A, {Schulreich} M~M, {Kinoshita} N,
  {Paul} M, {Dettbarn} C, {Fifield} L~K, {Golser} R, {Honda} M, {Linnemann} U,
  {Matsuzaki} H, {Merchel} S, {Rugel} G, {Steier} P, {Tims} S~G, {Winkler} S~R
  and {Yamagata} T 2017 {The Link Between the Local Bubble and Radioisotopic
  Signatures on Earth} in {\em 14th International Symposium on Nuclei in the
  Cosmos (NIC2016)\/} p 010304 (\textit{Preprint} \eprint{1611.01431})

\bibitem{Feige2017b}
{Schulreich} M~M, {Breitschwerdt} D, {Feige} J and {Dettbarn} C 2017 {Numerical
  studies on the link between radioisotopic signatures on Earth and the
  formation of the Local Bubble. I. $^{60}$Fe transport to the solar system by
  turbulent mixing of ejecta from nearby supernovae into a locally homogeneous
  interstellar medium} {\em \aap\/} {\bf 604} A81 (\textit{Preprint}
  \eprint{1704.08221})

\end{thebibliography}
\providecommand{\newblock}{}

\end{document}